\documentclass[twocolumn]{article}
\let\ifapj\iffalse
\let\ifarxiv\iftrue
\let\iflocal\iffalse
\usepackage{etoolbox,ifxetex,ifluatex}
\ifxetex\else\ifluatex\else\usepackage[utf8]{inputenc}\fi\fi
\ifluatex\AtEndPreamble{\hypersetup{pdfencoding=auto}}\fi
\usepackage{csquotes,siunitx}
\ifapj
  \usepackage{acro,appendix,hyperref}
  
  \let\symsf\mathsf
  \let\symcal\mathcal
\fi
\makeatletter\ifboolexpr{bool{arxiv} or bool{local}}{
  \usepackage{font-termes,draft}
  \usepackage[physics, astro]{basic}
  \usepackage[mode=biblatex, colorlinks]{apjbib}
  \ExecuteBibliographyOptions{embedlinks}
  \addbibresource{mri.bib}
  \renewcommand*\incr@eqnum
    {\refstepcounter{equation}\ifmmode\mathopen\fi{}\let\incr@eqnum\@empty}
  \hypersetup{urlcolor=blue}
}{}\makeatother
\DeclareAcronym{MHD}{
  short           =MHD,
  long            =magnetohydrodynamic,
  short-indefinite=an,
  short-plural    =,
  long-plural     =s}
\DeclareAcronym{GRMHD}{
  short           =GRMHD,
  long            =general-relativistic magnetohydrodynamic,
  short-plural    =,
  long-plural     =s}
\DeclareAcronym{MRI}{
  short           =MRI,
  long            =magnetorotational instability,
  short-indefinite=an}
\DeclareAcronym{TDE}{
  short           =TDE,
  long            =tidal disruption event}
\DeclareAcronym{ISCO}{
  short           =ISCO,
  long            =innermost stable circular orbit,
  short-indefinite=an,
  long-indefinite =an}
\newcommand*\astF{\mathord\ast F}
\expandafter\def\csname editcolor1\endcsname{blue!60!cyan}
\ifapj\let\edit\relax\fi
\newcommand*\edit[2]{\textcolor{\csname editcolor#1\endcsname}{#2}}

\usepackage{ulem,xcolor}
\newif\ifshowedit\showedittrue
\newcommand*\DeclareEditCommands[2]{
  \expandafter\newcommand\expandafter*\csname sig#1\endcsname
    {\textsf{\bfseries #1}}
  \expandafter\newcommand\csname add#1\endcsname[1]{%
    \ifshowedit
      \begingroup\color{#2}\csname sig#1\endcsname\ ##1\endgroup
    \else
      ##1%
    \fi}
  \expandafter\newcommand\csname del#1\endcsname[1]{%
    \ifshowedit
      \begingroup\color{#2}\csname sig#1\endcsname\ \sout{##1}\endgroup
    \fi}
  \expandafter\newcommand\csname rep#1\endcsname[2]{%
    \ifshowedit
      \begingroup\color{#2}\csname sig#1\endcsname\ \sout{##1} ##2\endgroup
    \else
      ##2%
    \fi}
  \expandafter\newcommand\csname note#1\endcsname[1]{%
    \ifshowedit
      \begingroup\color{#2}\guillemotleft##1 ---%
        \csname sig#1\endcsname\guillemotright\endgroup
    \fi}}
\DeclareEditCommands{TP}{red}
\DeclareEditCommands{JK}{blue}
\DeclareEditCommands{EC}{green!70!black}
\begin{document}

\title{Nonlinear evolution of the magnetorotational instability in eccentric
disks}

\ifapj
  \author[0000-0001-5949-6109]{Chi-Ho Chan}
  \affiliation{Center for Relativistic Astrophysics and School of Physics,
  Georgia Institute of Technology, Atlanta, GA 30332, USA}
  \author[0000-0002-7964-5420]{Tsvi Piran}
  \affiliation{Racah Institute of Physics, Hebrew University of Jerusalem,
  Jerusalem 91904, Israel}
  \author[0000-0002-2995-7717]{Julian~H. Krolik}
  \affiliation{Department of Physics and Astronomy, Johns Hopkins University,
  Baltimore, MD 21218, USA}
\fi

\ifboolexpr{bool{arxiv} or bool{local}}{
  \author[1]{Chi-Ho Chan}
  \author[2]{Tsvi Piran}
  \author[3]{Julian~H. Krolik}
  \affil[1]{Center for Relativistic Astrophysics and School of Physics,
  Georgia Institute of Technology, Atlanta, GA 30332, USA}
  \affil[2]{Racah Institute of Physics, Hebrew University of Jerusalem,
  Jerusalem 91904, Israel}
  \affil[3]{Department of Physics and Astronomy, Johns Hopkins University,
  Baltimore, MD 21218, USA}
}{}

\date{May 11, 2022}
\ifapj\else\uats{%
  Magnetohydrodynamical simulations (1966);
  Accretion (14);
  Black hole physics (159);
  Gravitation (661)}
\fi

\shorttitle{Nonlinear eccentric MRI}
\shortauthors{Chan et al.}
\pdftitle{Nonlinear evolution of the magnetorotational instability in eccentric
disks}
\pdfauthors{Chi-Ho Chan, Tsvi Piran, Julian H. Krolik}

\begin{abstract}
The \ac{MRI} has been extensively studied in circular magnetized disks, and its
ability to drive accretion has been demonstrated in a multitude of scenarios.
There are reasons to expect eccentric magnetized disks to also exist, but the
behavior of the \ac{MRI} in these disks remains largely uncharted territory.
Here we present the first simulations that follow the nonlinear development of
the \ac{MRI} in eccentric disks. We find that the \ac{MRI} in eccentric disks
resembles circular disks in two ways, in the overall level of saturation and in
the dependence of the detailed saturated state on magnetic topology. However,
in contrast with circular disks, the Maxwell stress in eccentric disks can be
negative in some disk sectors, even though the integrated stress is always
positive. The angular momentum flux raises the eccentricity of the inner parts
of the disk and diminishes the same of the outer parts. Because material
accreting onto a black hole from an eccentric orbit possesses more energy than
material tracing the innermost stable circular orbit, the radiative efficiency
of eccentric disks may be significantly lower than circular disks. This may
resolve the \textquote{inverse energy problem} seen in many \aclp{TDE}.
\end{abstract}
\acresetall

\section{Introduction}

Eccentric gaseous disks arise in a surprisingly wide variety of astrophysical
contexts. A number of mechanisms can explain the existence of eccentric disks,
the most commonly invoked one being external perturbation. In eccentric
binaries, secular gravitational interaction endows forced and free
eccentricities upon circumbinary and circumobject disks
\citep[e.g.,][]{2000ssd..book.....M}; in circular binaries, tidal forces couple
to circumobject disks through the $3:1$ mean motion resonance and allow free
eccentricity to grow exponentially \citep{1991ApJ...381..259L}. Another
possibility is that disks become more eccentric over time. Viscous
overstability \citep{1978MNRAS.185..629K}, which amplifies small-scale
eccentric perturbations in isolated disks \citep[e.g.,][]{1994MNRAS.266..583L,
2001MNRAS.325..231O}, is often cited in this connection. A third option is for
disks to be born eccentric. Outgassing from planetesimals can create eccentric
disks \citep{2021MNRAS.505L..21T}, and so can the tidal disruption of stars
\citep{2015ApJ...804...85S, 2015ApJ...806..164P, 2017MNRAS.467.1426S} and
molecular clouds \citep[e.g.,][]{2008Sci...321.1060B} by supermassive black
holes. On the phenomenological side, eccentric disks are sometimes invoked to
explain asymmetric lines in white dwarfs \citep[e.g.,][]{2006Sci...314.1908G},
as well as asymmetric broad emission lines in active galactic nuclei
\citep[e.g.,][]{1995ApJ...438..610E, 2021MNRAS.506.6014T} and \acp{TDE}
\citep[e.g.,][]{2014ApJ...783...23G, 2017MNRAS.472L..99L}.

Because ideal \acp{MHD} is supported even by low levels of ionization
\citep{1994ApJ...421..163B, 1996ApJ...457..355G}, we expect magnetic fields to
play a role in many of the eccentric disks enumerated above. The presence of
magnetic fields changes the way disks evolve because of the \ac{MRI}
\citep{1991ApJ...376..214B, 1991ApJ...376..223H}. Simply put, in a disk whose
inner parts rotate faster than the outer parts, differential rotation can latch
onto horizontal bits of the magnetic field, stretch them out, and amplify them.
The gas connected to one of these bits on the inside is pulled back by magnetic
tension, loses angular momentum, migrates inward, and picks up orbital speed.
In the meantime, the gas on the outside is dragged forward, gains angular
momentum, drifts outward, and slows down. The rising velocity difference across
the horizontal magnetic field in turn enhances its stretching, precipitating an
instability.

Analytic calculations for circular disks show that a perturbation can grow by
orders of magnitude per orbit in the linear stage \citep{1991ApJ...376..214B},
making the \ac{MRI} among the most vigorous \ac{MHD} instabilities. The
initially exponential amplification eventually enters the nonlinear stage and
breaks down into \ac{MHD} turbulence. Orbital shear enforces a correlation
between the radial and azimuthal components of the turbulent velocity, and
between the same components of the magnetic field. Turbulent stresses transport
angular momentum outward; gas robbed of angular momentum sinks to smaller
radii, and the disk accretes.

The saturation process is amenable only to numerical investigation. Previous
simulations of circular disks, numbering in the hundreds, are divided into
shearing-box simulations, which consider a small neighborhood of the disk as
representative of the whole \citep[e.g.,][]{1992ApJ...400..595H,
1995ApJ...440..742H, 1996ApJ...464..690H, 1995ApJ...446..741B,
1996ApJ...463..656S}, and global simulations, which take in the entire disk
\citep[e.g.,][]{1991ApJ...376..223H, 1998ApJ...501L.189A, 1999ASSL..240..195M,
2000ApJ...528..462H, 2003ApJ...592.1060D}. This large body of work converged on
the consensus that, irrespective of the circumstances simulated, \ac{MHD}
turbulence in circular disks saturates within 10 to 20 orbits, and the stresses
at saturation correspond to a \citet{1973A&A....24..337S} alpha parameter
between \numrange[range-phrase={ and }]{0.01}{1}. For circular disks around
black holes, the alpha parameter may change substantially near their inner
edges at the \ac{ISCO}. In disks with gas-dominated pressure, the alpha
parameter can increase very rapidly as matter approaches and crosses the
\ac{ISCO} \citep{2010ApJ...711..959N}; alternatively, in super-Eddington
radiation-dominated disks, it may exhibit a sharp peak at a radius a short
distance outside the \ac{ISCO} \citep{2014ApJ...796..106J}.

There is no reason to expect the \ac{MRI} and the associated \ac{MHD} stresses
to be absent from eccentric disks, though their character and the exact manner
in which they approach saturation may differ from circular disks. Very little
heed, however, has hitherto been paid to any aspect of the role of magnetic
fields in eccentric disks. We were the first to establish analytically that
eccentric disks are susceptible to the \ac{MRI} \citep{2018ApJ...856...12C}.
Compared to the circular \ac{MRI}, the growth rate of the eccentric \ac{MRI} is
smaller at the order-unity level and the range of unstable wavelengths is
wider. That work, however, is incomplete because it examined only linear
stability. It remains an open question whether the robust growth of \ac{MHD}
stresses in the linear stage would, as the \ac{MRI} turns nonlinear, translate
to saturated stress levels significant enough to affect disk evolution.

As of writing, only a couple of simulations have looked at how eccentricity
interacts nonlinearly with the \ac{MRI}. \Citet{2020MNRAS.497..451D} set up a
circular disk in the potential of \citet{1980A&A....88...23P} and excited
eccentric waves from the outer edge; they saw that the \ac{MRI} is active at
large radii where eccentricities are higher, but suppressed near the \ac{ISCO}
where eccentricities are lower and differential apsidal precession is stronger.
\Citet{2021MNRAS.505....1O} fed the disk around one member of a binary through
Roche-lobe overflow; they found that the $r\phi$\nobreakdash-component of the
magnetic stress aids tidal gravity in growing eccentricity but all other stress
components oppose it. Neither work achieved eccentricities
$\mathrelp\gtrsim0.1$ because doing so requires overcoming two challenges.

The first problem stems from the fact that existing Newtonian \acp{MHD} codes
are capable of handling only Cartesian, cylindrical, and spherical coordinate
systems. If a moderately eccentric disk were simulated in one of these
coordinate systems, the streamlines would be oblique to the grid, and so would
the magnetic fields dragged out by orbital shear. Numerical artifacts
reflecting grid symmetry would creep in; at the same time, excessive numerical
dissipation would prevent the disk from maintaining its shape over a long time.
These drawbacks limited previous simulations of isolated eccentric disks to
eccentricities small enough to be implementable as an $m=1$ perturbation to the
initial velocity \citep[e.g.,][]{2005A&A...432..757P}, but alternative
approaches do exist \citep{2016MNRAS.458.3739B, 2020MNRAS.497..435D,
2020MNRAS.497..451D}. Here we demonstrate that Newtonian simulations can be
performed in arbitrary coordinate systems with \acp{GRMHD} codes, provided that
the metric is judiciously chosen. Employing a coordinate system molded to the
shape of moderately eccentric disks significantly suppresses the numerical
errors arising from ordinary cylindrical coordinates.

The second difficulty is with the simulation setup. One may think that the
setting of localized perturbations in \citet{2018ApJ...856...12C} lends itself
naturally to shearing-box simulations \citep[e.g.,][]{2014MNRAS.445.2621O,
2018MNRAS.477.4838W}. Drawing inspiration from circular disks, one may imagine
shearing boxes in eccentric disks to have edges running along curves of
constant semilatus rectum and constant azimuth. However, the very notion of an
eccentric shearing box is suspect. Circular shearing boxes assume that the
disk, being homogeneous, is equivalent to a tiling of the shearing box; this
justifies periodic azimuthal and shift-periodic radial boundary conditions. The
assumption breaks down for eccentric disks. \citet{2018ApJ...856...12C} showed
that a perturbation grows differently at different positions along the orbit,
depending on the local orbital shear. This means the conditions at the leading
edge of an eccentric shearing box are different from the trailing edge, and
they also vary along each of the other two edges, so boundary conditions that
directly copy one edge to the other would be inappropriate. We can avoid these
questions about the eccentric shearing box by performing global simulations
instead. It is worth noting that the first simulations of the circular \ac{MRI}
were also global \citep{1991ApJ...376..223H}.

We recount our simulation setup in \cref{sec:methods}. The results from the
simulations are presented in \cref{sec:results} (see
\href{https://youtube.com/playlist?list=PLlhsZldWhMs6OIvfFxf5DZy9UbpimeGGt}{movies})
and discussed in \cref{sec:discussion}. Our concluding remarks are gathered in
\cref{sec:conclusions}.

\section{Methods}
\label{sec:methods}

We outline our simulation strategy in \cref{sec:overall strategy}. We continue
with the details of the simulation setup in the subsequent subsections and in
the Appendix; readers uninterested in the technicalities may skip to the
results in \cref{sec:results}.

\subsection{Overall strategy}
\label{sec:overall strategy}

Our goal is to simulate the nonlinear evolution of the eccentric \ac{MRI} in a
purely Newtonian setting. The only reason we turn to a \acp{GRMHD} code is
because numerical issues force us to tailor the coordinate system to the
eccentric disk shape, but existing Newtonian codes lack the facility to deal
with bespoke coordinate systems. The \acp{GRMHD} code we use for this purpose
is Athena++ \citep{2016ApJS..225...22W, 2020ApJS..249....4S}.

There can be drawbacks to solving Newtonian problems with a \acp{GRMHD} code,
principally the large truncation error potentially created by the smallness of
the typical kinetic and internal energies compared to the rest energy. This
error can, however, be mitigated by careful design of the simulation setup, as
described in later subsections. With our setup, the truncation error is
\num{\lesssim e-7} for the vast majority of cells.

A major difference of \acp{GRMHD} codes compared with Newtonian codes is that
gravity enters not as explicit momentum and energy source terms, but through
the metric. Our choice of the metric in \cref{sec:coordinates,sec:potential}
ensures that orbits are closed ellipses that do not apsidally precess, allowing
our simulations to closely approximate Newtonian behavior.

We simulate a suite of disks. The initial hydrodynamic configuration of the
disks follows the common prescription in \cref{sec:hydrodynamic initial
condition}: Gas is placed within a limited radial range, so that the inner and
outer edges of the disk are well-separated from the inner and outer boundaries,
respectively, of the simulation domain. All disks are tracked for 15 orbits so
turbulence may have enough time to reach saturation. The individual disks
comprising the suite are designed with contrast in mind. They are classified
along two orthogonal dimensions: circular versus eccentric, unmagnetized versus
magnetized.

Comparison between circular and eccentric disks gives us an idea whether the
saturation level of the \ac{MRI} depends on eccentricity. The eccentric disks
have a moderate eccentricity of 0.5, so that the character of the \ac{MRI}
specific to eccentric disks can reveal itself without being overwhelmed by
hydrodynamic effects.

Comparison between unmagnetized and magnetized disks helps us disentangle
\ac{MHD} effects from hydrodynamic effects. Magnetized disks are seeded with an
initial magnetic field as described in \cref{sec:magnetic initial condition}.
Two magnetic topologies are considered, vertical- and dipolar-field, because
topology can influence the saturated \ac{MHD} turbulence
\citep[e.g.,][]{1995ApJ...440..742H, 1996ApJ...464..690H, 2004ApJ...605..321S,
2013ApJ...767...30B}.

\subsection{Equations}

We employ natural units, which means the length and velocity units are the
gravitational radius and the speed of light, respectively. The sign convention
is $(-,+,+,+)$, Greek indices range over $\{0,1,2,3\}$, Latin indices range
over $\{1,2,3\}$, and Einstein summation is implied.

The equations of \acp{GRMHD} are
\begin{alignat}{3}
& \partial_t[(-g)^{1/2}\rho u^t
  &]& +\partial_j[(-g)^{1/2}\rho u^j
  &]&= 0, \\
& \partial_t[(-g)^{1/2}T^t_\mu
  &]& +\partial_j[(-g)^{1/2}T^j_\mu
  &]&= (-g)^{1/2}T^\nu_\sigma\Gamma^\sigma_{\mu\nu}, \\
& \partial_t[(-g)^{1/2}\astF^{it}
  &]& +\partial_j[(-g)^{1/2}\astF^{ij}
  &]&= 0.
\end{alignat}
Here $t$ is the coordinate time, $\rho$ is the comoving mass density, $u^\mu$
is the velocity, $g$ is the determinant of the metric $g_{\mu\nu}$, and
$\Gamma^\sigma_{\mu\nu}$ is the Christoffel symbol of the second kind. From the
Hodge dual of the electromagnetic tensor $\astF^{\mu\nu}$ we obtain the
magnetic field $B^i=\astF^{i0}$ and the projected magnetic field
$b^\mu=u_\nu\astF^{\nu\mu}$. Lastly, the stress--energy tensor is
\begin{equation}
T^{\mu\nu}=\biggl(p+\frac12b_\sigma b^\sigma\biggr)g^{\mu\nu}
  +\biggl(\rho+\frac\gamma{\gamma-1}p+b_\sigma b^\sigma\biggr)u^\mu u^\nu
  -b^\mu b^\nu,
\end{equation}
where $p$ and $\gamma$ are the gas pressure and adiabatic index, respectively.

\subsection{Eccentric coordinate system}
\label{sec:coordinates}

We solve the \acp{GRMHD} equations in an eccentric coordinate system
\citep{2001MNRAS.325..231O}. It is similar to the cylindrical coordinate
system, except that circular coordinate surfaces are replaced by axially
aligned elliptical ones, chosen such that their cross sections along the
midplane match the initial disk streamlines. Adapting the coordinate system to
the disk reduces numerical artifacts and dissipation in our simulations. The
eccentricities and orientations of the coordinate surfaces can in principle
vary from one elliptical cylinder to the next, but here we specialize to the
case in which both are spatially uniform.

Let $(t,R,\varphi,z)$ be cylindrical coordinates. We work with gravity weak
enough to be well-described by a quasi-Newtonian potential $\Phi(R,z)$; the
nonzero components of the metric in this limit are
\begin{align}
g_{tt} &= -(1+2\Phi), \\
g_{RR} &= 1, \\
g_{\varphi\varphi} &= R^2, \\
g_{zz} &= 1.
\end{align}
Let $(t,\log\lambda,\phi,z)$ be eccentric coordinates specialized for use in
our simulations; they are related to cylindrical coordinates by
\begin{align}
R &= \lambda/(1+e\cos\phi), \\
\varphi &= \phi.
\end{align}
Here $e$ is the eccentricity of our eccentric coordinates, set to 0 for
circular disks and 0.5 for eccentric disks. Coordinate surfaces of constant
$\lambda$ are elliptical cylinders of semilatus rectum $\lambda$, or
equivalently, semimajor axis $a=\lambda/(1-e^2)$. We opt for $\log\lambda$
instead of $\lambda$ in order to generate a logarithmic grid, but for ease of
understanding we continue to label that coordinate by $\lambda$ and express
results in terms of $\lambda$. We reuse $t$ and $z$ without risk of ambiguity
because these two coordinates do not participate in the coordinate
transformation from cylindrical to eccentric. The metric and connection in both
coordinates are provided in \cref{sec:metric}.

\subsection{Gravitational potential and orbits}
\label{sec:potential}

We ignore vertical gravity in these first simulations of the eccentric
\ac{MRI}, so the potential depends only on $R$. In this sense, our simulations
resemble earlier simulations of unstratified circular disks
\citep[e.g.,][]{1991ApJ...376..223H, 1995ApJ...440..742H, 1996ApJ...464..690H,
2004ApJ...605..321S, 2007A&A...476.1113F, 2007A&A...476.1123F,
2007MNRAS.378.1471L, 2009ApJ...690..974S, 2009ApJ...694.1010G,
2011ApJ...739...82B}. However, instead of the point-mass gravitational
potential $\Phi(R,z)=-1/R$, we adopt
\begin{equation}
\Phi(R,z)=-1/(R+2)
\end{equation}
because, as proven in \cref{sec:potential derivation}, this potential admits
closed eccentric orbits at all distances. The modification matters because
general-relativistic apsidal precession, albeit small at large distances,
accumulates over the tens of orbits during which the \ac{MRI} saturates.

The properties of orbits in our potential are also derived in
\cref{sec:potential derivation}; here we repeat the parts that support our
exposition. For an orbit along a coordinate curve of semilatus rectum $\lambda$
or semimajor axis $a=\lambda/(1-e^2)$, the orbital period is
\begin{equation}\label{eq:orbital period}
T=2\pi a^{3/2}(1+2/a).
\end{equation}
The specific energy and angular momentum conserved with respect to our metric
are
\begin{align}
\label{eq:orbit energy}
E &= (1+1/a)^{-1/2}, \\
\label{eq:orbit angular momentum}
L &= \lambda^{1/2}E.
\end{align}
The specific energy includes the rest energy; $E=1$ corresponds to marginally
bound material, and $E\to0$ as material becomes more bound. The nonzero
components of the orbital velocity are
\begin{align}
\label{eq:orbit velocity 0} u^t &= E/[1+2\Phi(R,z)], \\
\label{eq:orbit velocity 2} u^\phi &= L/R^2,
\end{align}
so the physical velocity at pericenter is
\begin{equation}
v_{\su p}(\lambda)=\frac{Ru^\phi}{u^t}=
  \frac{\lambda^{1/2}(1+e)}{\lambda+2(1+e)}.
\end{equation}

Consider a collection of such orbits nested within one another, all following
coordinate curves. The velocity field thus generated has finite divergence:
\begin{equation}
(-g)^{-1/2}\partial_j[(-g)^{1/2}u^j]=
  \frac{u^\phi}{R+2}\frac{e\sin\phi}{1+e\cos\phi}.
\end{equation}
Consequently, if the motion of a gas without pressure is described by this
velocity field, the density along a streamline of constant $\lambda$ cannot be
uniform, but must instead vary with $\phi$ in proportion to
\begin{equation}
d(\lambda,\phi)\eqdef
  \biggl[\frac{\lambda+2(1+e\cos\phi)}{\lambda+2(1+e)}\biggr]^{1/2}.
\end{equation}
So as to guarantee an initial condition that is a genuine hydrodynamical steady
state, we take this modulation into account in \cref{sec:hydrodynamic initial
condition} even though the modulation amplitude is tiny for our disk
parameters.

\subsection{Hydrodynamic initial condition}
\label{sec:hydrodynamic initial condition}

Because we ignore vertical gravity, our initial disk is translationally
symmetric along the $z$\nobreakdash-direction. It can be described as an
elliptical annular cylinder, each shell of which orbits the coordinate axis
along a coordinate surface of constant $\lambda$ with a velocity as given by
\cref{eq:orbit velocity 0,eq:orbit velocity 2}. The scale of the disk is
characterized by its fiducial orbit, whose semilatus rectum is the geometric
mean of the semilatera recta of its inner and outer edges.

The disk should be large enough that orbital velocities are non-relativistic,
and small enough that severe truncation errors do not arise from the evolution
of the total energy as a result of the smallness of the kinetic energy with
respect to the rest energy. In light of the fact that the linear growth rate of
the \ac{MRI} is inversely proportional to the orbital period
\citep{2018ApJ...856...12C}, we additionally require that runs of different $e$
have the same semimajor axis and thus orbital period. We settle on a semilatus
rectum of $\lambda_*=200(1-e^2)$ for the fiducial orbit, and we report time in
units of the orbital period at this orbit.

The initial density profile is
\begin{equation}\label{eq:initial density}
\rho(\lambda,\phi)=\rho_*m(\lambda,\phi),
\end{equation}
with $\rho_*$ the density at the pericenter of the fiducial orbit,
$(\lambda,\phi)=(\lambda_*,0)$. To give the disk edges that are not too sharp
and to build in a numerical vacuum, the density is modulated spatially as
\begin{equation}
m(\lambda,\phi)=
  (1-f_{\su v})d(\lambda,\phi)h(l_{\su{gb}},l_{\su{gt}};2q^\lambda)+f_{\su v},
\end{equation}
where
\begin{equation}
h(a,s;x)\eqdef
  \frac14\biggl(1+\tanh\frac{x+a}s\biggr)\biggl(1+\tanh\frac{a-x}s\biggr)
\end{equation}
is a smoothed top-hat function and
\begin{equation}
q^\lambda\eqdef
  \frac{\log\lambda-\log\lambda_*}{\log\lambda_{\su{max}}-\log\lambda_{\su{min}}}
\end{equation}
is the fractional $(\log\lambda)$\nobreakdash-position within the simulation
domain. The modulation is governed by three parameters: $f_{\su v}=0.01$ is the
vacuum-to-disk density ratio, and $l_{\su{gb}}=0.5$ and $l_{\su{gt}}=0.1$ are
the fractional $(\log\lambda)$\nobreakdash-extents of the disk body and the
disk--vacuum transition, respectively.

Because the most unstable mode of the circular \ac{MRI} is incompressible, the
thermodynamic properties of the gas are expected to have little bearing on the
growth of the eccentric \ac{MRI}. Even so, we would like the pressure to be
small enough initially that the configuration above is close to equilibrium,
and the internal energy to be large enough at all times that catastrophic
truncation errors are avoided. We balance these competing desires by setting
the Mach number at the fiducial pericenter to $M_*=30$. Additionally, we adopt
an adiabatic index of $\gamma=1+10^{-5}$, corresponding to a nearly isothermal
gas, so that the internal energy is larger at fixed pressure. The initial
pressure is therefore
\begin{equation}
p_*=\rho_*[v_{\su p}(\lambda_*)/M_*]^2/\gamma
\end{equation}
at the fiducial pericenter and
\begin{equation}
p(\lambda,\phi)=p_*[m(\lambda,\phi)]^\gamma
\end{equation}
everywhere else. This initial condition is not strictly hydrostatic; transient
outgoing waves are launched from the inner edge as the disk seeks force
balance. Moreover, our use of a soft equation of state means that density
perturbations are stronger and pressure gradients have a lesser effect on disk
evolution than if an adiabatic equation of state were used.

\subsection{Magnetic initial condition}
\label{sec:magnetic initial condition}

\begin{figure}
\includegraphics{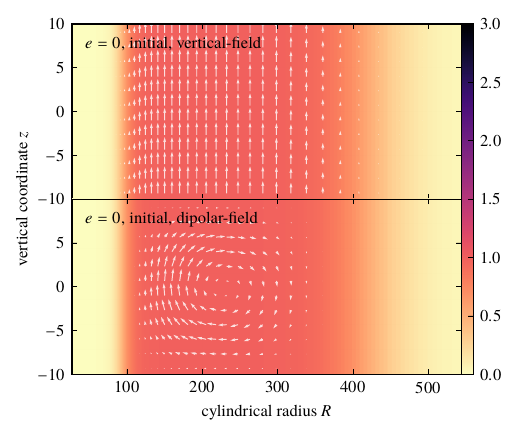}
\caption{Poloidal slices showing with arrows the initial magnetic field for our
two magnetic topologies. Background colors plot the initial density on the same
scale as \cref{fig:density}. The ordinate is more stretched than the abscissa,
and the vectors are arbitrarily scaled.}
\label{fig:topology}
\end{figure}

The magnetized runs are initialized with the two kinds of magnetic topologies
illustrated in \cref{fig:topology}. The vertical-field topology refers to a
magnetic field with one nonzero component
\begin{equation}
B^z\propto d(\lambda,\phi)h(l_{\su{mb}},l_{\su{mt}},2q^\lambda),
\end{equation}
where $l_{\su{mb}}=0.4$ and $l_{\su{mt}}=0.1$ are the fractional
$(\log\lambda)$\nobreakdash-extents of the magnetized disk body and the
transition from the magnetized disk body to the unmagnetized disk edges,
respectively. The net vertical magnetic flux persists throughout the
simulation. Thanks to its simplicity and its ability to generate the
fastest-growing instabilities, the vertical-field topology was considered in
the first studies of the circular \ac{MRI} \citep{1991ApJ...376..214B,
1991ApJ...376..223H} and in our analytic study of the eccentric \ac{MRI}
\citep{2018ApJ...856...12C}.

The dipolar-field topology is derived from a magnetic potential with one
nonzero component
\begin{equation}
A_\phi\propto
  (-g)^{1/2}\cos^2(\min(\tfrac12,\abs{q^\lambda/l_{\su{mb}}})\pi)\cos^2(q^z\pi),
\end{equation}
where
\begin{equation}
q^z\eqdef\frac z{z_{\su{max}}-z_{\su{min}}}
\end{equation}
is the fractional $z$\nobreakdash-position within the simulation domain and
$l_{\su{mb}}=0.4$. The inclusion of the metric determinant makes the
magnetic-field strength more uniform over azimuth. The dipolar-field topology
sees frequent application in global simulations of the circular \ac{MRI}
\citep[e.g.,][]{2000ApJ...528..462H}.

For both topologies, the initial plasma beta, defined as the ratio of the
initial volume integral of gas pressure to that of magnetic pressure, is 100.
The pressure due to the magnetic field is subtracted from the gas to preserve
the total. The gas pressure in magnetized regions is perturbed at the 0.01
level to seed the \ac{MRI}.

\subsection{Simulation domain, boundary conditions, and other numerical
concerns}

The simulation domain spans
$[\exp(-2)\lambda_*,\exp(1)\lambda_*]\times[-\pi,\pi]\times[-10,10]$ in
$(\lambda,\phi,z)$. The lower end of the $\lambda$\nobreakdash-range ensures
velocities are never close to the speed of light, and the asymmetry of the
$\lambda$\nobreakdash-range gives the infalling disk more room to evolve freely
before hitting the inner boundary. The resolutions of the unmagnetized and
magnetized simulations are $240\times240\times1$ and $240\times240\times60$,
respectively.

Periodic boundary conditions apply to the $\phi$\nobreakdash-direction. They
are also employed in the $z$\nobreakdash-direction, in accordance with our
neglect of vertical gravity. Outflow boundary conditions are used in the
$\lambda$\nobreakdash-direction: We copy all quantities to the ghost zone, zero
the $\lambda$\nobreakdash-component of the velocity if it points into the
simulation domain, and zero the $\phi$- and $z$\nobreakdash-components of the
magnetic field always. This last step reduces unphysical influences from the
boundaries.

To prevent numerical issues, we require that the pressure in the simulation
domain always satisfy $(\gamma p/\rho)^{1/2}\ge\num{2e-4}$. In addition,
whenever the recovery of primitive variables fails, the primitive variables
from the previous time step are carried forward.

\begin{figure}
\includegraphics{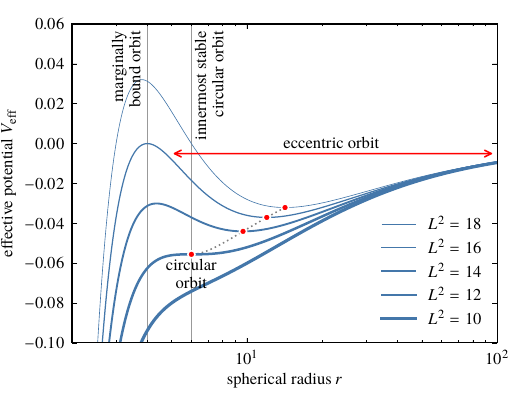}
\caption{Effective potential $V_{\su{eff}}=-1/r+\tfrac12(L^2/r^2)(1-2/r)$ in
Schwarzschild spacetime as a function of spherical radius $r$ and specific
angular momentum $L$. The specific energy $E$ must satisfy
$\tfrac12(E^2-1)>V_{\su{eff}}$, and marginally bound trajectories have $E=0$.
Material accretes by losing angular momentum; thus, its trajectory is described
by potentials of decreasing $L$. In circular disks, trajectories have the
lowest energy allowed by the stable potential well; such trajectories evolve
along the sequence of dots downward until they arrive at the smallest radius
that supports circular orbits, the \ac{ISCO}. In eccentric disks, trajectories
have energies above the bottom of the potential well, which allows them to make
radial excursions as suggested by the double-headed arrow. The smallest radius
a bound trajectory can reach without falling in is the marginally bound orbit;
such a trajectory has $E=0$ and $L^2=16$. Because trajectories energetic enough
to overcome the centrifugal barrier have energies greater than \iac{ISCO}
orbit, material plunging into the black hole on these trajectories has less
energy available for radiation compared to material accreting on circular
trajectories.}
\label{fig:effective potential}
\end{figure}

As the simulation progresses, different parts of the disk may evolve
differently in eccentricity and orientation, so the disk could eventually
become misaligned with the grid, resulting in greater numerical dissipation.
The disk also occupies a wider range of semilatera recta due to pressure
gradients and outward angular momentum transport, bringing the now differently
shaped disk into contact with the boundaries. The inner boundary poses a lower
limit on the pericenter, but this restriction is arguably physical because
there are indeed radii an accretion flow cannot return from. If the central
object is a star, the disk cannot extend inside the star or its magnetosphere.
If the central object is a black hole, \cref{fig:effective potential} informs
us that material with sufficiently low angular momentum and high energy can
plunge directly into the black hole; low angular momentum and high energy are,
of course, the hallmarks of an eccentric orbit. By contrast, the interaction
with the outer boundary is unphysical and can lead to numerical artifacts;
therefore, we restrict our attention to the first 15 orbits at $a=200$, before
the disk starts running into the outer boundary and numerical artifacts appear.
Steady state appears to obtain at this time for the plasma beta and the alpha
parameter, despite the continual evolution in disk shape.

\section{Results}
\label{sec:results}

\subsection{Overview}

\begin{figure}
\includegraphics{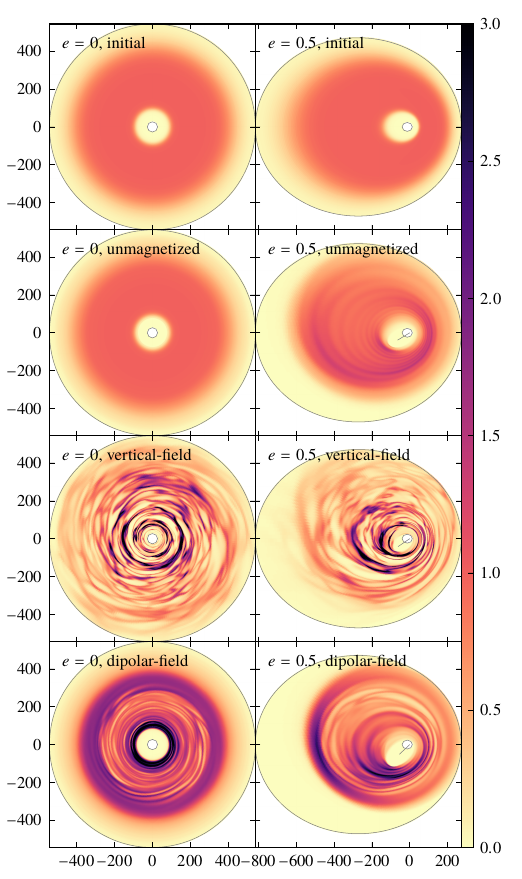}
\caption{Midplane slices of density, in units of the fiducial density $\rho_*$
from \cref{eq:initial density}. The top row shows two initial disks with
different eccentricities. The panels under each top-row panel are the outcomes
of imposing various magnetic topologies on an initial disk and evolving it for
15 orbits. The boundaries of the simulation domain are traced by thin ellipses
in order to better distinguish low-density regions from regions not covered by
the simulations. Short lines from the origin indicate the approximate
orientations of the inner parts of the eccentric disks.}
\label{fig:density}
\end{figure}

\Cref{fig:density} tells us how much the disks have changed by the end of the
simulations. The circular unmagnetized disk after 15 orbits is almost
indistinguishable from the initial disk. The eccentric unmagnetized disk
changes shape slightly because pressure gradients, which are nonzero along the
inner and outer disk edges, cause differential apsidal precession
\citep[e.g.,][]{2001MNRAS.325..231O}; the density striation is a result of this
adjustment process.

The most conspicuous contrast between unmagnetized and magnetized disks, of
whatever eccentricity, is that unmagnetized disks remain smooth while
magnetized disks develop large density fluctuations. This, of course, is due to
the \ac{MRI} creating \ac{MHD} turbulence in the magnetized disks. The
fluctuations are larger in vertical-field disks than in dipolar-field disks.

Comparison can also be made between circular and eccentric disks regardless of
magnetic topology. Unlike the circular disks, which stay circular despite the
\ac{MHD} turbulence, the inner parts of eccentric disks grow more eccentric and
precess prograde. As the inner parts of eccentric disks shrink, their
pericenters move inside the inner boundary, and their material accretes across
the boundary while retaining its eccentricity. This loss of material from the
most eccentric orbits is the reason why there is a sparsely populated region
between the inner parts and the inner boundary, visible in the late-time
eccentric disks in \cref{fig:density}.

There are also differences among the eccentric disks, concerning chiefly
eccentricity evolution and to a lesser degree precession. The unmagnetized and
vertical-field disks in \cref{fig:density} have largely preserved their initial
eccentricity, even though the outer parts of the unmagnetized disk have become
somewhat rounder, and the inner parts of the vertical-field disk have precessed
slightly more. The dipolar-field disk features the steepest eccentricity
gradient and a much higher degree of precession.

\begin{figure}
\includegraphics{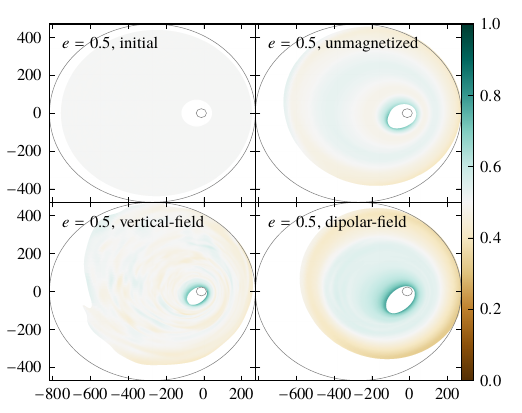}
\caption{Mass-weighted vertical average of the instantaneous eccentricity, as
defined in \cref{eq:instantaneous eccentricity map}. The top-left panel shows
the eccentric initial disk. The other panels are the outcomes of imposing
various magnetic topologies on the initial disk and evolving it for 15 orbits.
The boundaries of the simulation domain are traced by thin ellipses. In keeping
with \cref{fig:angular momentum squared and binding energy}, regions with
density below $0.1\,\rho_*$ are left blank, where $\rho_*$ is the fiducial
density from \cref{eq:initial density}.}
\label{fig:instantaneous eccentricity}
\end{figure}

The eccentricity gradient can be quantified by computing the instantaneous
eccentricity $\bar e$, which is the eccentricity of the orbit material would
follow given its instantaneous velocity $u^\mu$ if only gravitational forces
act; this orbit is also known as the osculating orbit. We calculate $\bar e$
from $u^\mu$ using \begingroup\csname@cref@sortfalse\endcsname\cref{eq:orbit
velocity 0,eq:orbit velocity 2,eq:eccentricity}\endgroup.
\Cref{fig:instantaneous eccentricity} contains plots of the mass-weighted
vertical average of $\bar e$:
\begin{equation}\label{eq:instantaneous eccentricity map}
\mean{\bar e}_{z;\rho}\eqdef\int dz\,\rho\bar e\bigg/\int dz\,\rho.
\end{equation}
The instantaneous eccentricity deviates little from its initial value in the
unmagnetized and vertical-field disks, but it develops a clear gradient in the
dipolar-field disk, with the eccentricity higher than its initial value in the
inner parts and lower in the outer parts. The implications of the eccentricity
gradient will be discussed in \cref{sec:inner edge}.

\subsection{Plasma beta and the alpha parameter}

\begin{figure}
\includegraphics{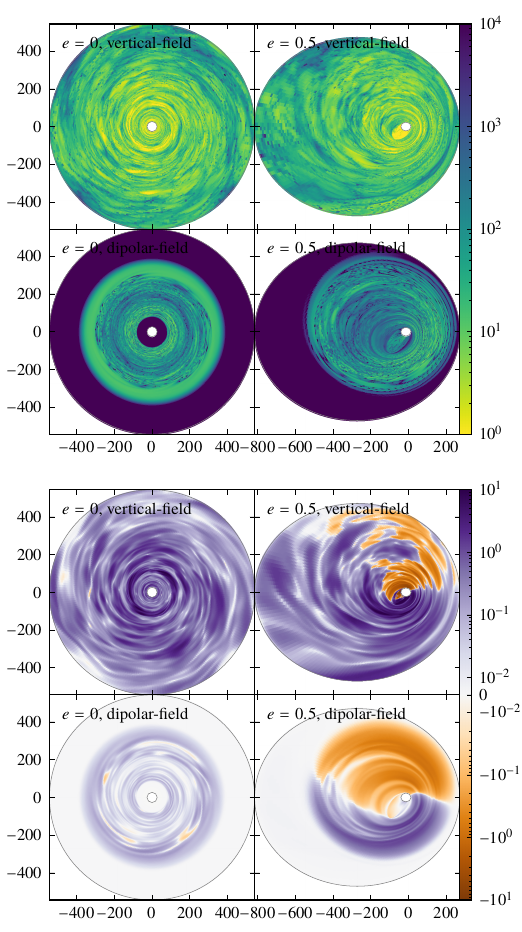}
\caption{Mass-weighted vertical averages of the plasma beta in the top half and
Maxwell-only alpha parameter in the bottom half, as defined in \cref{eq:plasma
beta map,eq:alpha parameter map}, respectively. The panels are the outcomes of
imposing various magnetic topologies on an initial disk and evolving it for 15
orbits. Regions with negligible levels of magnetic field have exceedingly large
values of plasma beta. The boundaries of the simulation domain are traced by
thin ellipses.}
\label{fig:magnetic field map}
\end{figure}

The top half of \cref{fig:magnetic field map} portrays the mass-weighted
vertical average of the plasma beta at the end of the simulations, defined as
\begin{equation}\label{eq:plasma beta map}
\mean\beta_{z;\rho}\eqdef
  \int dz\,\rho\frac{2p}{b_\mu b^\mu}\bigg/\int dz\,\rho.
\end{equation}
The plasma beta is \num{\sim10} in vertical-field disks and \num{\sim100} in
dipolar-field disks; the variation within a disk is about one order of
magnitude. Comparable levels of plasma beta are witnessed in circular and
eccentric disks with the same magnetic topology, suggesting that the \ac{MRI}
is unimpeded by eccentricity, in contrast to the findings of
\citet{2020MNRAS.497..451D}. The stronger magnetic fields produced by the
vertical-field topology accord with simulations of circular disks in the
literature \citep[e.g.,][]{1995ApJ...440..742H, 1996ApJ...464..690H,
2004ApJ...605..321S, 2013ApJ...767...30B}.

\begin{figure}
\includegraphics{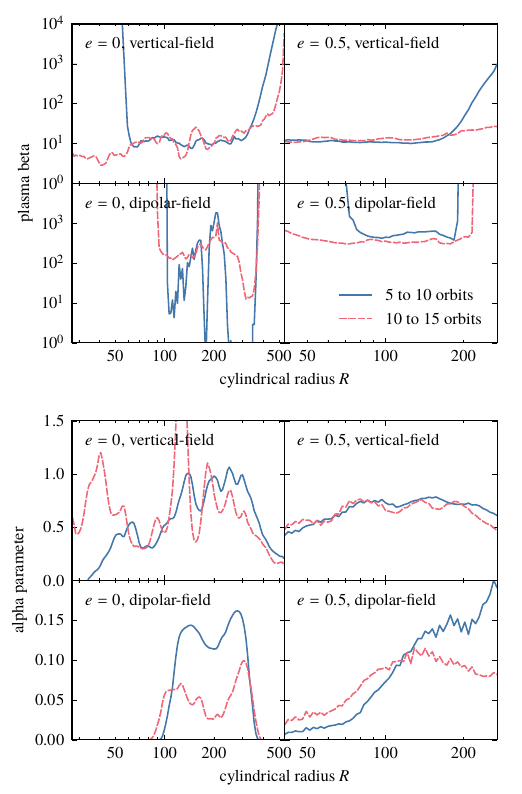}
\caption{Mass-weighted radial profiles of the plasma beta in the top half and
Maxwell-only alpha parameter in the bottom half, as defined in \cref{eq:plasma
beta profile,eq:alpha parameter profile}, respectively. The profiles are
time-averaged over the two intervals in the legend. The two rows in the bottom
half have different vertical scales. The magnetic field strength varies weakly
with eccentricity, but is much stronger in vertical-field disks than in
dipolar-field disks.}
\label{fig:magnetic field profile}
\end{figure}

We can also examine the plasma beta along a one-dimensional profile running
from the inside of the disk to the outside at different times during the
simulations. Considering that the disk evolves in eccentricity and orientation,
it makes little sense to look at profiles over semilatus rectum; instead, we
construct profiles over cylindrical radius. The mass-weighted average plasma
beta at cylindrical radius $R$ is given by
\begin{equation}\label{eq:plasma beta profile}
\mean\beta_{t\varphi z;\rho}\eqdef
  \int dS\,\rho\frac{2p}{b_\mu b^\mu}\bigg/\int dS\,\rho,
\end{equation}
where the hypersurface element is
\begin{equation}\label{eq:hypersurface element}
dS\eqdef(-g)^{1/2}\,dt\,d(\log\lambda)\,d\phi\,dz
  \mathop\delta\biggl(\frac\lambda{1+e\cos\phi}-R\biggr).
\end{equation}
Temporal and spatial averaging smooths out turbulent fluctuations. Temporal
averaging is performed over two intervals, each lasting one-third of the
simulation duration; comparison between the intervals gives us an idea how
close the \ac{MRI} is to saturation. Spatial averaging is limited to the
cylindrical shell of radius $R$ picked out by the delta-function. The results
are plotted in the top half of \cref{fig:magnetic field profile}, and the
legend lists the intervals of temporal averaging. The relatively small
difference between the two intervals at radii $100\lesssim R\lesssim200$
suggests that steady state is achieved to some degree at those radii, despite
the relatively short simulation duration. We also reach similar conclusions as
we did with \cref{fig:magnetic field map}: The plasma beta is quite uniform
over the disk, it is not significantly modified by the introduction of
eccentricity, but it is one to two orders of magnitude lower in vertical-field
disks than in dipolar-field disks.

The alpha parameter is conventionally taken to be the sum of Reynolds and
Maxwell stresses divided by the gas pressure. However, it is difficult to
determine the mean flow and departures from it in a disk whose inner and outer
parts evolve differently in eccentricity and orientation. We therefore consider
only the Maxwell, not Reynolds, contribution to the alpha parameter, working
under the assumption that the Maxwell stress dominates the total, as is
uniformly the case for circular disks \citep[e.g.,][]{1995ApJ...440..742H}. The
Maxwell stress is defined in terms of the projected magnetic field $b^\mu$,
similar to the stress--energy tensor. To make the stress more physically
interpretable, we measure $b^\mu$ in a local orthonormal basis, whose basis
vectors are those of cylindrical coordinates but with lengths normalized to
unity:
\begin{align}
\label{eq:physical magnetic field 1}
b^{\hat R}
  &= R\biggl(b^\lambda+b^\phi\frac{e\sin\phi}{1+e\cos\phi}\biggr), \\
\label{eq:physical magnetic field 2}
b^{\hat\varphi}
  &= Rb^\phi, \\
\label{eq:physical magnetic field 3}
b^{\hat z}
  &= b^z.
\end{align}
We then define the Maxwell stress to be $-b^{\hat\varphi}b^{\hat R}$. A factor
of $R$ is attached to $b^\lambda$ in \cref{eq:physical magnetic field 1}
because our eccentric coordinates use $\log\lambda$, not $\lambda$.
Orthonormality guarantees that
\begin{equation}
g_{\lambda\lambda}b^\lambda b^\lambda+2g_{\lambda\phi}b^\lambda b^\phi
  +g_{\phi\phi}b^\phi b^\phi=
  b^{\hat R}b^{\hat R}+b^{\hat\varphi}b^{\hat\varphi}.
\end{equation}

The bottom half of \cref{fig:magnetic field map} depicts the mass-weighted
vertical average of the Maxwell-only alpha parameter at the end of the
simulations:
\begin{equation}\label{eq:alpha parameter map}
\mean{\alpha_{\su m}}_{z;p}\eqdef
  -\int dz\,b^{\hat\varphi}b^{\hat R}\bigg/\int dz\,p.
\end{equation}
The alpha parameter is positive almost everywhere in circular disks, as
required for outward angular momentum transport. By contrast, the alpha
parameter is uniformly positive in the lower half of the eccentric disks where
material falls to pericenter, but it is consistently \textit{negative} in
certain sectors of the upper half where material flies out to apocenter. In the
eccentric vertical-field disk, positive sectors occupy significantly more area
than negative sectors. The magnitude of the alpha parameter, whether positive
or negative, varies from \numrange{\sim0.2}{5}; however, if we construct
area-weighted histograms of the magnitude separately for positive and negative
sectors, we find that the positive histogram is shifted by a factor of
\num{\sim2} toward larger values relative to its negative counterpart. In the
eccentric dipolar-field disk, the total area of positive sectors is only
slightly larger than that of negative sectors. In addition, the magnitude of
the alpha parameter has a narrower distribution, \numrange{\sim0.3}{3}; the
positive histogram is again displaced by a factor of \num{\sim2} compared to
the negative one. We shall speculate about why the alpha parameter switches
sign in \cref{sec:stresses}.

The net effect of angular momentum transport is revealed by integrating the
Maxwell stress over an orbit. When doing so by eye on the basis of
\cref{fig:magnetic field map}, it is important to take into account the
relative areas and alpha-parameter ranges of the positive and negative sectors.
More quantitatively, we construct in \cref{fig:magnetic field profile}
mass-weighted radial profiles of the alpha parameter at different times using
the prescription
\begin{equation}\label{eq:alpha parameter profile}
\mean{\alpha_{\su m}}_{t\varphi z;p}\eqdef
  -\int dS\,b^{\hat\varphi}b^{\hat R}\bigg/\int dS\,p,
\end{equation}
with $dS$ the same hypersurface element from \cref{eq:hypersurface element}.
The alpha parameter is comparable in circular and eccentric disks with the same
magnetic topology. In vertical-field disks, the alpha parameter is
\numrange{\sim0.5}{1}; in dipolar-field disks, it is still positive, but only
\numrange{\sim0.05}{0.15}. Stronger stress for vertical than dipolar magnetic
field agrees with previous simulations of circular disks
\citep[e.g.,][]{1995ApJ...440..742H, 1996ApJ...464..690H, 2004ApJ...605..321S,
2013ApJ...767...30B}.

\subsection{Specific angular momentum and binding energy}
\label{sec:angular momentum and binding energy}

To investigate the effect of internal stresses, we examine how the specific
angular momentum squared $L^2$ and binding energy $E_{\su b}=1-E$ evolve. The
mass-weighted vertical averages of the two quantities at the end of the
simulations are
\begin{align}
\label{eq:angular momentum squared map}
\mean{L^2}_{z;\rho} &\eqdef
  \int dz\,\rho(Ru^{\hat\varphi})^2\bigg/\int dz\,\rho, \\
\label{eq:binding energy map}
\mean{E_{\su b}}_{z;\rho} &\eqdef
  \int dz\,\rho(1+u_t)\bigg/\int dz\,\rho.
\end{align}
Here $u^{\hat\varphi}=Ru^\phi$ is the velocity measured in the local
orthonormal cylindrical basis, defined analogously to $b^{\hat\varphi}$ in
\cref{eq:physical magnetic field 2}, and the covariant velocity component
$u_t=-E$ is a conserved quantity of our time-independent metric.

\begin{figure*}
\includegraphics{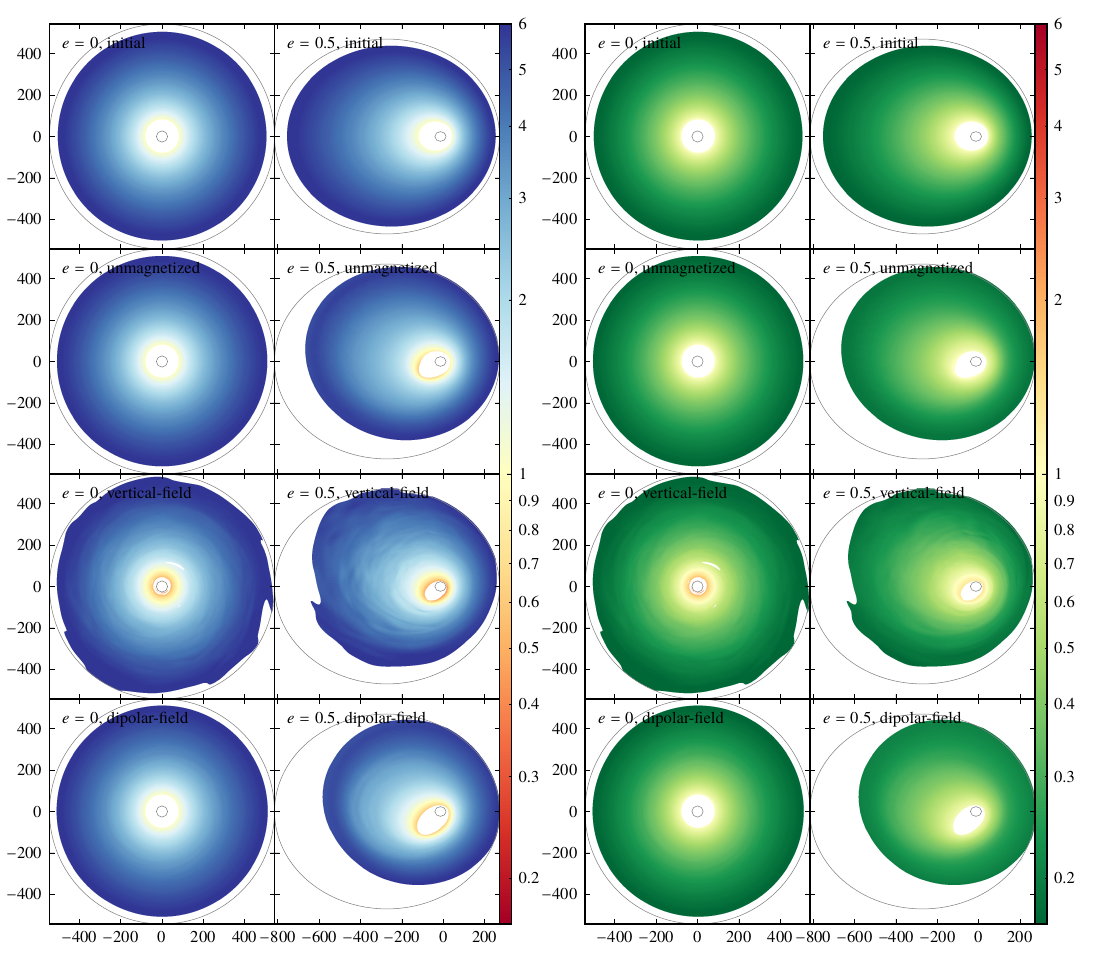}
\caption{Mass-weighted vertical averages of the specific angular momentum
squared in the left half and specific binding energy in the right half, as
defined in \cref{eq:angular momentum squared map,eq:binding energy map},
respectively. The top row shows two initial disks with different
eccentricities. The panels under each top-row panel are the outcomes of
imposing various magnetic topologies on an initial disk and evolving it for 15
orbits. In all panels, the inner edge is arbitrarily defined to be where the
density is $0.1\,\rho_*$, with $\rho_*$ the fiducial density from
\cref{eq:initial density}, and regions less dense than that are left blank.
Furthermore, the specific angular momentum squared or binding energy in each
panel is normalized by its value at the inner edge of the corresponding initial
disk, which is why the inner edge appears yellow in the top row. The boundaries
of the simulation domain are traced by thin ellipses. The fractional changes in
the specific angular momentum squared and binding energy together determine the
change in eccentricity seen in \cref{fig:instantaneous eccentricity}.}
\label{fig:angular momentum squared and binding energy}
\end{figure*}

\Cref{fig:angular momentum squared and binding energy} displays the specific
angular momentum squared and binding energy, normalized to their initial values
at the inner edge. Our focus is on the eccentric disks. Because of pressure
gradients, even the inner edge of the unmagnetized disk experiences a reduction
in specific angular momentum squared by \SI{\sim25}{\percent} from its initial
value. The decrease is greater in magnetized disks where the Maxwell stress
also contributes: The dipolar-field disk reports a drop by
\SI{\sim33}{\percent}, and the vertical-field disk, which has a smaller plasma
beta and larger alpha parameter than the dipolar-field disk, records a
suppression by \SI{\sim50}{\percent}. It should be noted that the range of
specific angular momentum is constrained by the geometry of the simulation
domain: Once material loses enough angular momentum that its pericenter recedes
inside the inner boundary, it leaves the simulation domain.

The specific binding energy at the inner edge changes by rather less. It is
almost identical to its initial value in the unmagnetized disk,
\SI{\sim30}{\percent} higher in the vertical-field disk, and
\SI{\sim10}{\percent} \textit{lower} in the dipolar-field disk. The smaller
fractional changes suggest that the torques transporting angular momentum
outward, regardless of whether they are hydrodynamic or magnetic in nature,
occur preferentially near apocenter where the attendant work done is smaller.

\subsection{Quality factors}

We close this section by examining how well our magnetized disks resolve the
\ac{MRI}. The figures of merit for circular disks are the quality factors,
defined as the ratio of the characteristic wavelength of the \ac{MRI} to cell
sizes in different directions \citep{2011ApJ...738...84H}. We generalize their
mass-weighted vertical averages to eccentric disks as
\begin{align}
\label{eq:quality 3}
\mean{Q_z}_{z;\rho} &\eqdef \int dz\,\rho\frac{\abs{b^{\hat z}}T}
  {\rho^{1/2}\Delta z}\bigg/\int dz\,\rho, \\
\label{eq:quality 2}
\mean{Q_\phi}_{z;\rho} &\eqdef \int dz\,\rho\frac{\abs{b^{\hat\varphi}}T}
  {\rho^{1/2}R\Delta\phi}\bigg/\int dz\,\rho,
\end{align}
where $T$ is the orbital period from \cref{eq:orbital period}, and $\Delta z$
and $\Delta\phi$ are the cell sizes in the $z$- and
$\phi$\nobreakdash-directions, respectively. In keeping with work on circular
disks, we base our quality factors on the physical, cylindrical components
$b^{\hat\mu}$ of the projected magnetic field $b^\mu$, given by
\cref{eq:physical magnetic field 2,eq:physical magnetic field 3}.

\begin{figure}
\includegraphics{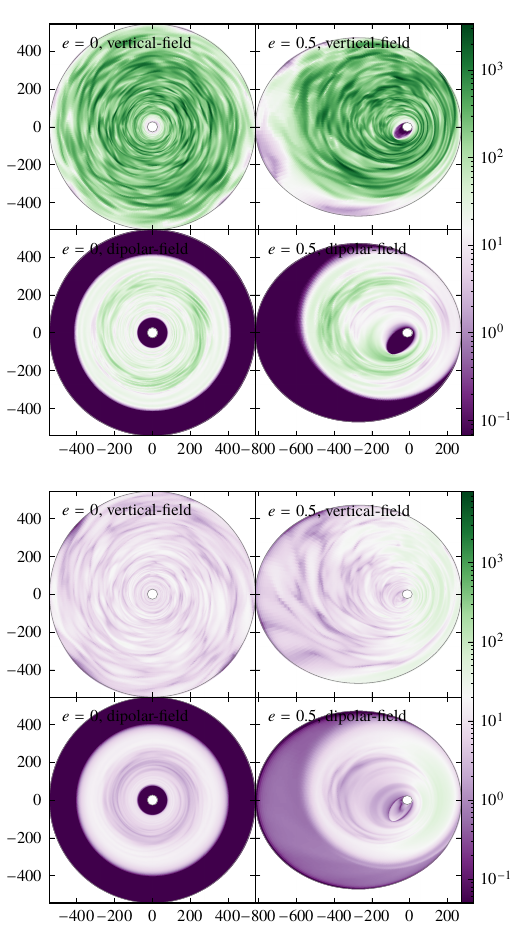}
\caption{Mass-weighted vertical averages of the vertical quality factor in the
top half and azimuthal quality factor in the bottom half, as defined in
\cref{eq:quality 3,eq:quality 2}, respectively. The panels are the outcomes of
imposing various magnetic topologies on an initial disk and evolving it for 15
orbits. The boundaries of the simulation domain are traced by thin ellipses.}
\label{fig:quality}
\end{figure}

\Cref{fig:quality} showcases the quality factors at the end of the simulations.
For ease of comparison with the criteria that indicate adequate resolution for
circular disks, to wit, $\mean{Q_z}\gtrsim15$ and $\mean{Q_\phi}\gtrsim20$
\citep{2013ApJ...772..102H}, the color scales of the figure are centered on
these values. In terms of $\mean{Q_z}$, the vertical-field disks are extremely
well-resolved everywhere, but the same is true for the dipolar-field disks only
for a limited range of semilatera recta. In terms of $\mean{Q_\phi}$, all disks
are marginally under-resolved. Interestingly, $\mean{Q_\phi}$ is noticeably
higher in the pericentric half of the disks; this is because the strong orbital
shear near pericenter amplifies $b^{\hat\varphi}$, not $b^{\hat z}$.

\section{Discussion}
\label{sec:discussion}

\subsection{Stresses in circular and eccentric disks}
\label{sec:stresses}

Our earlier investigation into the linear stage of the \ac{MRI} found that the
\ac{MRI} grows in both circular and eccentric disks; the growth rate in
eccentric disks is about half that in circular disks
\citep{2018ApJ...856...12C}. The present simulations suggest that the nonlinear
stage of the \ac{MRI} is also not so different in circular and eccentric disks,
in that it saturates to comparable levels of plasma beta and alpha parameter in
the two kinds of disks. Thus, the \ac{MRI} functions in much the same capacity
in eccentric disks as in circular disks, mediating outward angular momentum
transport.

That being said, the \ac{MRI} in eccentric disks exhibits two intriguing
features not witnessed in circular disks. The first one is the sign flip in the
Maxwell-only alpha parameter. In circular disks, the alpha parameter is
positive almost everywhere; by contrast, in eccentric disks, the alpha
parameter can be consistently negative in some sectors of the disk even though
the azimuthal integral of the Maxwell stress remains positive.

To understand the basic principle underlying this surprising behavior, we break
temporarily from the general-relativistic treatment used in the rest of the
article and work in the Newtonian limit. We denote the Newtonian velocity and
magnetic field by $\vec{\tilde v}$ and $\vec{\tilde B}$, respectively, and
their components measured against the local orthonormal cylindrical basis by
$\tilde v_i$ and $\tilde B_i$. We conflate $\tilde B_i$ with $b^{\hat i}$
elsewhere in the text, so the Maxwell stress is simply $-\tilde B_R\tilde
B_\varphi$. Our starting point is the induction equation:
\begin{equation}
\pd{\vec{\tilde B}}t=\curl(\vec{\tilde v}\cross\vec{\tilde B}).
\end{equation}
To track the time evolution of the magnetic field at a point comoving with the
flow, we consider
\begin{equation}
\od{\vec{\tilde B}}t
  =\pd{\vec{\tilde B}}t+(\vec{\tilde v}\cdot\grad)\vec{\tilde B}
  =(\vec{\tilde B}\cdot\grad)\vec{\tilde v}
  -\vec{\tilde B}(\divg\vec{\tilde v}).
\end{equation}
We set $\divg\vec{\tilde v}=0$ because the \ac{MRI} is largely incompressible,
and we take $\pds{}z=0$ because we ignore vertical gravity in our simulations.
Consequently,
\begin{equation}\label{eq:Maxwell stress evolution}
\od{}t(-\tilde B_R\tilde B_\varphi)=
  -\tilde B_R^2\pd{\tilde v_\varphi}R
  -\tilde B_\varphi^2\biggl(\frac1R\pd{\tilde v_R}{\varphi}\biggr)
  -(-\tilde B_R\tilde B_\varphi)\frac{\tilde v_R}R.
\end{equation}

In circular disks, the time- and azimuth-averaged $\pds{\tilde v_\varphi}R$ is
negative, the averaged $(1/R)(\pds{\tilde v_R}\varphi)$ is zero, and the
averaged $\tilde v_R$ is also negative, albeit with a very small magnitude.
Because $-\tilde B_R\tilde B_\varphi>0$ nearly everywhere, all terms on the
right-hand side of \cref{eq:Maxwell stress evolution} are therefore on average
zero or positive, so $-\tilde B_R\tilde B_\varphi$ grows over time until
limited by dissipation.

In eccentric disks, the time-averaged $\pds{\tilde v_\varphi}R$ is also
negative everywhere. However, the averaged $(1/R)(\pds{\tilde v_R}\varphi)$ is
no longer zero: It changes from negative to positive shortly before pericenter
and back after pericenter. The averaged $\tilde v_R$ also changes sign, from
negative to positive at pericenter and back at apocenter. The second and third
terms therefore have azimuth-dependent signs. When orbits are significantly
eccentric, the averaged $(1/R)(\pds{\tilde v_R}\varphi)$ is comparable in
magnitude to the averaged $\pds{\tilde v_\varphi}R$, and the averaged $\tilde
v_R$ is comparable to the averaged $\tilde v_\varphi$, so all three terms can
be important.

Consider material starting from apocenter with positive Maxwell stress, that
is, $-\tilde B_R\tilde B_\varphi>0$. Initially all three terms work together to
make $-\tilde B_R\tilde B_\varphi$ more positive. As the material swings toward
pericenter and its velocity becomes more azimuthal than radial,
$(1/R)(\pds{\tilde v_R}\varphi)$ turns positive while $\abs{\tilde v_R}$ drops,
flipping the sign of the second term and reducing the magnitude of the third
term. If $\abs{\tilde B_R}\lesssim\abs{\tilde B_\varphi}$, which is marginally
satisfied at these azimuths, the second term can dominate, pulling $-\tilde
B_R\tilde B_\varphi$ toward negative values. For some distance past pericenter,
the second term continues to be negative; at the same time, $\tilde v_R>0$ and
$\ods{\tilde v_R}t>0$, so the third term now tends to reduce the magnitude of
$-\tilde B_R\tilde B_\varphi$ at an increasing rate regardless of its sign. The
second and third terms combined make $-\tilde B_R\tilde B_\varphi<0$ at some
point near pericenter. Further beyond pericenter, the second term changes back
to positive and the third term decreases in magnitude. By the time the material
returns to apocenter, the first and second terms have restored $-\tilde
B_R\tilde B_\varphi$ to positive, transporting angular momentum outward.

Both magnetic topologies share these qualitative features, but they differ in
the quantitative details. The dipolar-field topology yields a very close
alignment between the local directions of velocity and magnetic field, whereas
the vertical-field topology produces merely a correlation between the two
directions that has much more scatter. In addition, as we can see in
\cref{fig:magnetic field map}, the sectors of the dipolar-field disk with
$-\tilde B_R\tilde B_\phi<0$ coincide with the region where $\tilde v_R>0$,
whereas the vertical-field disk sectors with $-\tilde B_R\tilde B_\phi<0$ take
up only a small part of the $\tilde v_R>0$ region, and their boundary is much
more irregular. We speculate that both of these contrasts are due to the larger
amplitude of turbulence the \ac{MRI} drives in the presence of a vertical
magnetic field, an effect well-documented in circular disks
\citep[e.g.,][]{1995ApJ...440..742H, 1996ApJ...464..690H, 2004ApJ...605..321S,
2013ApJ...767...30B}. Because the third term of \cref{eq:Maxwell stress
evolution} is sensitive to correlations between fluctuations in velocity and
magnetic field, large-amplitude turbulence may change the stress evolution.
This supposition is corroborated by the fact that the vertical root-mean-square
of $u^z$, a good proxy for the magnitude of turbulent fluctuations in
unstratified simulations, in the vertical-field disk is \num{\sim4} times that
in the dipolar-field disk.

The other curiosity is that, whereas \ac{MHD} stresses in circular disks
transport angular momentum and energy at such rates as to keep shrinking orbits
circular, the two rates can be independent of each other in eccentric disks.
Circular disks have $E_{\su b}L^2\approx\tfrac12$ at all radii, according to
\cref{eq:Newtonian eccentricity}; by contrast, $E_{\su b}$ and $L^2$ in
eccentric disks vary by amounts that depend on location, and the mismatch
dictates how the local eccentricity evolves. The fractional changes of $E_{\su
b}$ and $L^2$ estimated in \cref{sec:angular momentum and binding energy} imply
that the unmagnetized disk has the least eccentric inner edge and the
dipolar-field disk has the most eccentric, which agrees with
\cref{fig:instantaneous eccentricity}. One possible explanation of the ranking
relies on the fact that stresses near pericenter tend to lower the eccentricity
of the inner edge and stresses near apocenter tend to raise it
\citep{2017MNRAS.467.1426S}. Stresses in the unmagnetized disk, being purely
hydrodynamical, should be concentrated near pericenter where pressure gradients
are the steepest; conversely, the Maxwell stress in the vertical- and
dipolar-field disks are relatively evenly distributed between pericenter and
apocenter, as made apparent by the bottom half of \cref{fig:magnetic field
map}. Our speculation in the previous paragraph may bear on the ordering of the
two magnetized disks: Stronger turbulence in the vertical-field disk disrupts
coherent angular momentum transport, so the inner edge is less eccentric than
in the dipolar-field disk.

\subsection{Dynamics and energy dissipation at the inner edge}
\label{sec:inner edge}

As demonstrated in \cref{sec:angular momentum and binding energy}, \ac{MHD}
stresses in eccentric disks are typically more effective at moving angular
momentum than energy. Because angular momentum constrains the size of the inner
edge, and because the binding energy at the inner edge determines the amount of
energy available for radiation, the different transport rates of these two
quantities could have observable effects on eccentric disks.

To explore these effects, let us first consider how disks behave around a star.
For a circular disk, the inner edge is located at the larger of the stellar or
Alfv\'en radius. The total energy a given amount of material dissipates in the
disk is its binding energy on a circular orbit at that radius. Additional
dissipation happens in the boundary layer at the inner edge as the material
comes into corotation with the star or its magnetosphere. If the star-regulated
rotation speed in this boundary layer is small compared to the orbital speed,
the additional dissipation is equal to the dissipation that took place in the
disk itself.

For an eccentric disk, the stellar or Alfv\'en radius likewise defines the
inner edge, but in this case, it is through a match to the pericenter of the
inner edge. Because the corresponding semimajor axis is larger than the stellar
or Alfv\'en radius, material dissipates less energy in the disk proper. The
total energy dissipated, however, is exactly the same as in the circular case
because the eccentric orbit of the material ultimately transforms into a
circular one at the stellar or Alfv\'en radius. What makes eccentric disks
different is that dissipation in the boundary layer accounts for a larger
fraction of the total. Energy can be dissipated there in a variety of ways,
depending on how the inner edge interacts with the star or its magnetosphere.
In one extreme, the material immediately comes into corotation. The fast-moving
material dissipates large amounts of energy in a small region, potentially
producing hard radiation. In the other extreme, the material loses a tiny part
of its kinetic energy each time it grazes past the star or its magnetosphere,
and the material migrates inward gradually on orbits of decreasing semimajor
axes and eccentricities. Dissipation in this case could be more spatially
distributed, happening both in the grinding encounters and along the
circularizing orbits; if so, the eccentric disk may resemble a circular one in
appearance.

We now turn to disks around black holes. As shown in \cref{fig:effective
potential}, material on eccentric orbits can reach the event horizon even if it
has more angular momentum than that of \iac{ISCO} orbit provided that it has
enough energy to overcome the centrifugal barrier inside the \ac{ISCO} radius;
equivalently, it can do so if its eccentricity is high enough. Moreover, when
the angular momentum is at least that of \iac{ISCO} orbit, the peak of the
centrifugal barrier is always above the energy of \iac{ISCO} orbit.
Consequently, the amount of energy available for such material to radiate is
always smaller than the binding energy at the \ac{ISCO}, the conventional
estimate for radiative efficiency in circular disks \citep{1973blho.conf..343N,
1974ApJ...191..499P}. The energy available to eccentric material is even
smaller when we take into account extra energy extraction by the Maxwell stress
in circular disks \citep{1974ApJ...191..507T, 1999ApJ...515L..73K,
1999ApJ...522L..57G, 2002ApJ...573..754K, 2009ApJ...692..411N,
2016MNRAS.462..636A, 2016ApJ...819...48S, 2021ApJ...922..270K}. Unlike
eccentric disks around stars, the energy retained by the plunging material
cannot be recovered for radiation in some boundary layer.

The diminution of the radiative efficiency of eccentric disks around black
holes could be particularly relevant to \acp{TDE}, in which the energy radiated
is considerably below that expected from the accretion of a reasonable fraction
of a stellar mass of material onto a black hole through a conventional circular
disk \citep{2015ApJ...806..164P}. \Citet{2017MNRAS.467.1426S} suggested that
this \textquote{inverse energy problem} could be resolved by internal stresses
that transport angular momentum preferentially and cause the most bound debris
to plunge; our results provide quantitative support to that qualitative
argument.

\subsection{Effects of vertical gravity}

Our simulations of the nonlinear development of the \ac{MRI} in eccentric disks
ignored vertical gravity. We did so for two reasons. The first is to isolate
the effects of orbital-plane motions, the chief driver of the \ac{MRI}, when
they are azimuthally modulated in eccentric disks. The second is to produce a
baseline for comparison with future simulations that do include vertical
gravity.

Because vertical gravity is a form of tidal gravity, it is much weaker than
radial gravity as long as the disk is thin. For this reason, the modifications
it introduces are likely minor and unable to quench the \ac{MRI}. Here we
examine what these modifications may be.

One effect of vertical gravity that is discernible in circular disks carries
over to eccentric disks: It regulates the buoyant eviction of accumulated
magnetic flux that may determine the saturation level of the \ac{MRI}
\citep[e.g.,][]{1992MNRAS.259..604T, 2010ApJ...713...52D, 2011MNRAS.416..361B,
2011ApJ...732L..30H, 2015ApJ...809..118B, 2018ApJ...861...24H}.

The variation of vertical gravity around an eccentric orbit causes the disk to
expand in height as it travels from pericenter to apocenter and to collapse on
the way back. This \textquote{breathing} can have large amplitudes even for
mildly eccentric disks \citep{2014MNRAS.445.2621O, 2021MNRAS.500.4110L}. For
the thickest and most eccentric disks, breathing can even become non-adiabatic:
Extreme compression can create shocks that eject material vertically
\citep{2021ApJ...920..130R}.

Vertical gravity can also influence the eccentric \ac{MRI} more subtly, without
shocks. Vertical motion opens up more avenues of energy exchange between the
orbit and the magnetic field. Compression and expansion can alter the Alfv\'en
speed, hence the critical wavelength for stability against the \ac{MRI}, and it
can also modulate the wavelengths of advected perturbations. All these
variations are periodic, raising the possibility of instability through
parametric resonance \citep{2005A&A...432..743P}.

It bears repeating that the changes due to vertical gravity should be small for
thin, moderately eccentric disks, so we expect our results here to remain
generally valid.

\section{Conclusions}
\label{sec:conclusions}

In circular disks, the \ac{MRI} stirs up correlated \ac{MHD} turbulence,
turbulent stresses transport angular momentum outward, and the disk accretes.
Our simulations demonstrate that much the same process operates in eccentric
disks, and with comparable efficiency. Like circular disks, the quantitative
level of Maxwell stress achieved in eccentric disks depends on the magnetic
topology, but eccentric and circular disks with the same magnetic topology
reach comparable levels of plasma beta and alpha parameter.

Although the mass-weighted, disk-averaged Maxwell stress in an eccentric disk
produces an outward angular momentum flux similar in magnitude to that in a
circular disk with the same magnetic topology, the Maxwell stress in an
eccentric disk can cause \textit{inward} angular momentum transport in certain
disk sectors. This behavior is seen only in eccentric disks likely because the
radial velocity and its azimuthal gradient are both nonzero and have signs that
vary over azimuth.

By removing proportionately more angular momentum than energy from the inner
parts of the disk, \ac{MHD} stresses can promote accretion of highly eccentric
material. This material has higher energy than a circular orbit with the same
angular momentum, so the radiative efficiency of a disk around a black hole can
be suppressed relative to a circular disk. These findings corroborate earlier
suggestions about how the power output from eccentric disks may explain why the
total observed radiated energy in many \acp{TDE} is one to two orders of
magnitude lower than expected from circular accretion of a stellar mass of
material \citep{2017MNRAS.467.1426S}.

\begin{acknowledgments}
The authors thank Scott Noble for useful discussions, as well as Elliot Lynch,
Janosz Dewberry, and an anonymous referee for comments which helped improve the
clarity of the text. CHC and TP were supported by ERC Advanced Grant
\textquote{TReX}. CHC was additionally supported by NSF grant AST-1908042. JHK
was partially supported by NSF grants AST-1715032 and AST-2009260. The
simulations were performed on the Rockfish cluster at the Maryland Advanced
Research Computing Center (MARCC).
\end{acknowledgments}

\software{Athena++ \citep{2016ApJS..225...22W, 2020ApJS..249....4S}, NumPy
\citep{2020Natur.585..357H}, SymPy \citep{10.7717/peerj-cs.103}, Matplotlib
\citep{2007CSE.....9...90H}}

\begin{appendices}

\section{Metric components and Christoffel symbols}
\label{sec:metric}

In cylindrical coordinates $(t,R,\varphi,z)$, the nonzero metric components are
\begin{align}
g_{tt} &= -\symsf P, \\
g_{RR} &= 1, \\
g_{\varphi\varphi} &= R^2, \\
g_{zz} &= 1, \\
g^{tt} &= -1/\symsf P, \\
g^{RR} &= 1, \\
g^{\varphi\varphi} &= 1/R^2, \\
g^{zz} &= 1,
\end{align}
the metric determinant is
\begin{equation}
g=-R^2\symsf P,
\end{equation}
and the nonzero Christoffel symbols of the second kind are
\begin{align}
\Gamma^t_{tR}=\Gamma^t_{Rt}
  &= \partial_R\Phi/\symsf P, \\
\Gamma^t_{tz}=\Gamma^t_{zt}
  &= \partial_z\Phi/\symsf P, \\
\Gamma^R_{tt}
  &= \partial_R\Phi, \\
\Gamma^R_{\varphi\varphi}
  &= -R, \\
\Gamma^\varphi_{R\varphi}=\Gamma^\varphi_{\varphi R}
  &= 1/R, \\
\Gamma^z_{tt}
  &= \partial_z\Phi,
\end{align}
where
\begin{equation}
\symsf P\eqdef1+2\Phi(R,z).
\end{equation}

Performing a coordinate transformation to eccentric coordinates
$(t,\log\lambda,\phi,z)$, we find that the nonzero metric components are
\begin{align}
g_{tt} &= -\symsf P, \\
g_{\lambda\lambda} &= R^2, \\
g_{\lambda\phi}=g_{\phi\lambda} &= R^2\symsf Q, \\
g_{\phi\phi} &= R^2(1+\symsf Q^2), \\
g_{zz} &= 1, \\
g^{tt} &= -1/\symsf P, \\
g^{\lambda\lambda} &= (1+\symsf Q^2)/R^2, \\
g^{\lambda\phi}=g^{\phi\lambda} &= -\symsf Q/R^2, \\
g^{\phi\phi} &= 1/R^2, \\
g^{zz} &= 1,
\end{align}
the metric determinant is
\begin{equation}
g=-R^4\symsf P,
\end{equation}
and the nonzero Christoffel symbols of the second kind are
\begin{align}
\Gamma^t_{t\lambda}=\Gamma^t_{\lambda t}
  &= (\partial_R\Phi)R/\symsf P, \\
\Gamma^t_{t\phi}=\Gamma^t_{\phi t}
  &= (\partial_R\Phi)R\symsf Q/\symsf P, \\
\Gamma^t_{tz}=\Gamma^t_{zt}
  &= \partial_z\Phi/\symsf P, \\
\Gamma^\lambda_{tt}
  &= \partial_R\Phi/R, \\
\Gamma^\lambda_{\lambda\lambda}
  &= 1, \\
\Gamma^\lambda_{\phi\phi}
  &= -R/\lambda, \\
\Gamma^\phi_{\lambda\phi}=\Gamma^\phi_{\phi\lambda}
  &= 1, \\
\Gamma^\phi_{\phi\phi}
  &= 2\symsf Q, \\
\Gamma^z_{tt}
  &= \partial_z\Phi,
\end{align}
where
\begin{equation}
\symsf Q\eqdef e\sin\phi/(1+e\cos\phi).
\end{equation}

\section{Precession-free gravitational potential}
\label{sec:potential derivation}

Schwarzschild spacetime in general relativity produces prograde apsidal
precession, while an extended gravitating mass in Newtonian mechanics causes
retrograde apsidal precession. It is natural to ask if one can be made to
cancel the other exactly.

We work in cylindrical coordinates $(t,R,\varphi,z)$. The metric in the
weak-gravity limit is given in \cref{sec:metric}, but here we assume $\Phi$ is
not yet decided. The velocity of a particle is $u^\mu$; without loss of
generality we restrict the particle to the midplane, so $z=0$ and $u^z=0$.
Because $t$ and $\varphi$ are ignorable in the metric, we immediately have two
integrals of motion
\begin{align}
\label{eq:conserved energy}
E &= u^t\symsf P, \\
\label{eq:conserved angular momentum}
L &= R^2u^\varphi.
\end{align}
Velocity normalization requires
\begin{equation}\label{eq:velocity normalization}
(u^R)^2=-1+E^2/\symsf P-L^2/R^2.
\end{equation}
Dividing the equation by $R^4(u^\varphi)^2=L^2$, changing variable to
$\xi\eqdef1/R$, and differentiating with respect to $\xi$ yields
\begin{equation}\label{eq:inverse radius equation}
\odd\xi\varphi+\xi=-\frac{E^2}{2L^2\symsf P^2}\od{\symsf P}\xi.
\end{equation}
Closed eccentric orbits exist for all values of $E$ and $L$ if and only if
\begin{equation}
-\frac1{2\symsf P^2}\od{\symsf P}\xi=C_1,
\end{equation}
which has the solution
\begin{equation}
\symsf P=1/(2C_1\xi+C_2),
\end{equation}
where $C_1$ and $C_2$ are constants. We take $C_1=C_2=1$ so that $\symsf
P=1+2\Phi\approx1-2\xi$ for $\xi\ll1$, as befitting point-mass gravity. The
potential $\Phi=-1/(R+2)$ describes softened gravity, so there are no
coordinate or physical singularities.

Our next step is to determine the velocity $u^\mu$ at any point along an
eccentric orbit. The solution to \cref{eq:inverse radius equation} is
\begin{equation}\label{eq:orbit solution}
R=\bar\lambda/[1+\bar e\cos(\varphi+C_3)],
\end{equation}
where $\bar e$ and $C_3$ are constants and
\begin{equation}\label{eq:semilatus rectum}
\bar\lambda\eqdef L^2/E^2.
\end{equation}
The embellishments on $\bar e$ and $\bar\lambda$ serve to distinguish them from
$e$ and $\lambda$ defining our eccentric coordinates. \Cref{eq:orbit solution}
describes an ellipse of eccentricity $\bar e$, semilatus rectum $\bar\lambda$,
and semimajor axis $\bar a=\bar\lambda/(1-\bar e^2)$. We pick $C_3=0$ so that
the pericenter of the ellipse is at $\varphi=0$, and we fix $\bar e$ by solving
\cref{eq:velocity normalization} at pericenter, where $u^R=0$:
\begin{equation}\label{eq:eccentricity}
\bar e^2=1+(E^2-1)L^2/E^4.
\end{equation}
This expression reduces to its Newtonian equivalent
\begin{equation}\label{eq:Newtonian eccentricity}
\bar e^2\approx1-2E_{\su b}L^2
\end{equation}
when $\abs{E_{\su b}}=\abs{1-E}\ll1$. Once we know $\bar e$ and $\bar\lambda$
for an eccentric orbit, we can solve for $E$ and $L$ using \cref{eq:semilatus
rectum,eq:eccentricity}; the results are \cref{eq:orbit energy,eq:orbit angular
momentum}. These can then be substituted into \cref{eq:conserved
energy,eq:conserved angular momentum,eq:velocity normalization} to yield the
velocity $u^\mu$ in cylindrical coordinates. The velocity $u^\mu$ in eccentric
coordinates, given in \cref{eq:orbit velocity 0,eq:orbit velocity 2}, follows a
coordinate transformation.

The final missing piece is the analogue of Kepler's equation for our potential.
We introduce the eccentric anomaly $\symcal E$, defined by
\begin{equation}
R=\bar a(1-\bar e\cos\symcal E).
\end{equation}
Using a result from geometry,
\begin{equation}
(1-\bar e)^{1/2}\tan\tfrac12\varphi=
  (1+\bar e)^{1/2}\tan\tfrac12\symcal E,
\end{equation}
we can express $t$ along the orbit as a function of $\varphi$:
\begin{equation}\label{eq:time}
t=\int d\symcal E\,\od\varphi{\symcal E}\frac{u^t}{u^\varphi}
  =\bar a^{3/2}[(1+2/\bar a)\symcal E-\bar e\sin\symcal E].
\end{equation}
The integration constant is set to zero for simplicity. As $\symcal E$
increases by $2\pi$, $t$ also increases by an amount equal to the orbital
period:
\begin{equation}
T=2\pi\bar a^{3/2}(1+2/\bar a).
\end{equation}
Multiplying both sides of \cref{eq:time} by $2\pi/T$ furnishes us with the
analogue of Kepler's equation:
\begin{equation}
\symcal M=\symcal E-\frac{\bar e\sin\symcal E}{1+2/\bar a},
\end{equation}
where $\symcal M=2\pi t/T$ is the mean anomaly.

\end{appendices}

\ifapj\bibliography{mri}\fi
\ifboolexpr{bool{arxiv} or bool{local}}{\bibhang1.25em\printbibliography}{}

\end{document}